\documentclass[12pt]{article}
\usepackage{amsmath,amssymb}

\makeatletter
\newcommand{\eg}{e.g.,~}
\newcommand{\ie}{i.e.,~}

\@addtoreset{equation}{section}
\makeatother

\newcommand{\al}{\alpha}

\begin{document}

\thispagestyle{empty}

\setcounter{page}{0}

\mbox{}
\vspace{5mm}

\begin{center} {\bf \Large  A dynamical inconsistency of Ho\v{r}ava gravity}

\vspace{1.5cm}

Marc Henneaux$^{1,2}$, Axel Kleinschmidt$^{1}$ and  Gustavo Lucena G\'omez$^{1}$

\footnotesize
\vspace{1 cm}

${}^1${\em Universit\'e Libre de Bruxelles and International Solvay Institutes, ULB-Campus Plaine CP231, 1050 Brussels, Belgium}

\vspace{.2cm}

${}^2${\em Centro de Estudios Cient\'{\i}ficos (CECS), Casilla 1469, Valdivia, Chile}\\

\vspace{7mm}
{\tt\{henneaux,axel.kleinschmidt,glucenag\}@ulb.ac.be}
\end{center}

\vspace {12mm}

\centerline{\bf Abstract}
\vspace{.2cm}
\noindent
The dynamical consistency of the non-projectable version of Ho\v{r}ava gravity is investigated by focusing on the asymptotically flat case. It is argued that for generic solutions of the constraint equations the lapse must vanish asymptotically. We then consider particular values of the coupling constants for which the equations are tractable and in that case we prove that the lapse must vanish everywhere -- and not only at infinity. Put differently, the Hamiltonian constraints are generically all second-class. We then argue that the same feature holds for generic values of the couplings, thus revealing a physical inconsistency of the theory. In order to cure this pathology, one might want to introduce further constraints but the resulting theory would then lose much of the appeal of the original proposal by Ho\v{r}ava. We also show that there is no contradiction with the time reparametrization invariance of the action, as this invariance is shown to be a so-called ``trivial gauge symmetry" in Ho\v{r}ava gravity, hence with no associated first-class constraints.

\newpage

\section{Introduction}

Recently, Ho\v{r}ava proposed a candidate for a UV completion of Einstein theory of gravity in which full spacetime diffeomorphism invariance is abandoned and recovered only at large distances \cite{Horava:2008ih,Horava:2009uw}.  Based on appealing analogies with condensed matter physics and anisotropic scaling \`a la Lifschitz (see~\cite{Horava:2008ih,Horava:2009uw} and references therein), it has been proposed that this alternative to Einstein theory might provide a renormalizable UV completion of general relativity and therefore potentially yields a very attractive approach that is worth being explored.

There are two classes of Ho\v{r}ava theories.  One is the class of the so-called ``projectable" theories, in which the lapse is restricted to depend only on time.  The other one is the ``non-projectable" class, where the lapse is allowed to depend on both space and time. In the first case, there is only one integrated Hamiltonian constraint $\int \text{d}^3 x\, {\mathcal H}(x) = 0$.  In the second case, there is an infinity of Hamiltonian constraints ${\mathcal H}(x) = 0$, one at each space point, just as in general relativity. As we are interested in theories that reproduce Einstein gravity in the IR limit, with its full set of constraints, we shall consider in this paper only the non-projectable class of theories, although comparison with the projectable case will be made when it illustrates useful points.  We shall also allow for all terms compatible with formal renormalizability while keeping the lapse and shift functions as Lagrange multipliers. In particular, we will not use the ``detailed balance'' condition which has been shown to be problematic for example for standard black hole solutions~\cite{Lu:2009em} (see also \cite{Charmousis:2009em}).

We begin by showing that, in the asymptotically flat case, the lapse must asymptotically tend to zero, thus preventing any interesting dynamics. This, in itself, is already a serious drawback.

We then pursue the analysis for a special choice of coupling constants that yields more tractable equations.  The results obtained in that case are argued to also hold for general values of the couplings. The main result derived then is that the Hamiltonian constraints ${\mathcal H}(x) = 0$ are generically all second-class. Namely, they completely determine the lapse, which must then vanish everywhere and not only at infinity.  Put differently, {\em there is generically no first-class constraint among the Hamiltonian constraints}. The rank of the ``matrix" of the Poisson brackets $[{\mathcal H}(x), {\mathcal H}(x')]$ of the constraints is generically maximal and its corank is zero. This result might appear to be in contradiction with the known time reparametrization invariance of the theory, but we prove that this is not the case as time reparametrization invariance is in this instance a so-called ``trivial symmetry'' with no implication on the dynamics (see \eg \cite{MH90,Henneaux:1992ig}).

By ``generically", we mean ``at a generic point of the constraint surface defined by the Hamiltonian and momentum constraints". Indeed, the rank of the ``matrix" of the brackets $[{\mathcal H}(x), {\mathcal H}(x')]$ does depend on the location on the constraint surface (a situation already somewhat pathological in itself and excluded by Dirac in his theory of constrained systems \cite{Dirac:1950pj,Henneaux:1992ig}).  When we consider generic values of the couplings, ``generically" also means ``at a generic point in the space of couplings", often taken to be in an open subset of maximal dimension in the space of couplings.

The fact that the lapse must generically vanish everywhere  (in the assymptotically flat case) appears to be a serious blow to the theory in its original formulation.  One might try to rescue it by imposing further constraints but the resulting theory, even if mathematically consistent, would seem to depart sufficiently from general relativity so that it would cease to be a meaningful candidate for a UV completion of Einstein theory of gravity. We make comments along these lines in the conclusions.

Our paper extends previous works which already questioned the consistency of Ho\v{r}ava theory \cite{Li:2009bg,Blas:2009yd} but we make here a more complete analysis of the constraint equations and show that there is no contradiction with explicit invariances of the action.  Although we disagree with some aspects of earlier analyses, as we shall comment below, we agree with their final conclusion.  Namely, we confirm the inconsistency of Ho\v{r}ava theory in its original formulation, which could only be regained at the price of drastic modifications that would make it lose much of its appeal.

The structure of our paper is as follows. In section \ref{sec:dynamics}, we give the dynamics of Ho\v{r}ava gravity in Hamiltonian form and derive an equation for the lapse function expressing that the Hamiltonian constraints are preserved in time. This crucial equation is analysed in more detail in section \ref{sec:lapse} in the general case and in a specific and simpler model in section \ref{sec:tractable}. A seeming paradox between the action possessing time reparametrization invariance and the absence of any associated first-class constraint is resolved in section \ref{sec:paradox}. We conclude the main part of the paper with comments on the viability of Ho\v{r}ava gravity as a theory of gravitation. Several more technical details of some of our arguments have been relegated to appendices.

\section{Dynamics}
\label{sec:dynamics}

We describe the dynamics of Ho\v{r}ava's non-projectable class of theories in Hamiltonian form. The Hamiltonian data consist in:
\begin{itemize}
\item Canonical variables on phase space variables, $g_{ij}(x)$, $\pi^{ij}(x)$, with Poisson brackets
\begin{equation}
{}[g_{ij}(x), \pi^{mn}(y)]= (\delta^{m}_{i}\delta^{n}_{j}+\delta^{n}_{i}\delta^{m}_{j})\delta(x,y),
\end{equation}
where $x$ and $y$ are points on a spatial slice and $\pi^{ij}$ is the momentum conjugated to the spatial metric $g_{ij}$.
\item A constraint surface in phase space defined by constraints
\begin{align}
\label{hamconstraints}
 {\mathcal H}(x) &\approx 0   \quad\quad&& \hbox{ (``Hamiltonian constraint"),}\\
 {\mathcal H}_k(x) &\approx 0  \quad\quad&&\hbox{ (``momentum constraints")}
\end{align}
where weak equality $\approx$ means zero on the constraint surface, as usual.
\item Equations of motion generated by a Hamiltonian
\begin{equation}\label{ham}
H = \int \text{d}^3x \left( N(x) \, {\mathcal H}(x) + N^k(x) \, {\mathcal H}_k(x) \right)
\end{equation}
with lapse function $N(x,t)$ and shift vector $N^k(x,t)$.\footnote{Very often, the time dependence is not written explicitly.  We focus on the space dependence.  In that spirit, a ``constant" means a function of $t$ only.}
\end{itemize}
The action is
\begin{equation}
S[g_{ij},\pi^{ij},N,N^k] = \int \text{d}t \left[\left(\int \text{d}^3x \, \pi^{ij} \dot{g}_{ij}\right) - H \right]. \label{action0}
\end{equation}
The equations of motion follow by extremizing the action with respect to $g_{ij}(x)$ and  $\pi^{ij}(x)$ (dynamical equations of motion), as well as with respect to the lapse and the shift functions that serve as Lagrange multipliers for the constraints.

The form of the momentum constraints, which generate spatial diffeomorphisms, is universal and given by
\begin{equation}
{\mathcal H}_k = - 2 \nabla_i \pi^i_{\; \; k}
\end{equation}
where $\nabla$ stands for the spatial covariant derivative operator. Indices are lowered and raised with the spatial metric $g_{ij}$ and its inverse $g^{ij}$.

By contrast, the Hamiltonian constraints depend on various coupling constants and take the form:
\begin{equation}
{\mathcal H} = {\mathcal H}_1 + {\mathcal H}_2
\end{equation}
where ${\mathcal H}_1$ is the kinetic term\footnote{$\lambda$ is the parameter appearing in the modified DeWitt metric on the space of metrics and is expected to go to zero in the IR limit if general relativity is to be recovered at low energies (see \cite{Horava:2009uw} for the details).} (with $\pi = g_{ij}\pi^{ij}$),
\begin{equation}
{\mathcal H}_1 = \frac{1}{\sqrt{g}}\left(\pi^{ij} \pi_{ij} - \frac{\lambda}{3 \lambda - 1} \pi^2 \right)
\end{equation}
and ${\mathcal H}_2$ contains the potential terms with up to six derivatives of the spatial metric
\begin{equation}\label{H2}
{\mathcal H}_2 = \sqrt{g} \left( \sigma + \xi R + \eta R^2 + \zeta R^{ij}R_{ij} +\beta C_{ij}C^{ij}+\gamma R \triangle R + \ldots\right).
\end{equation}
Here, the spatial Laplacian is $\triangle = \nabla^i \nabla_i$. The restriction to six derivatives comes from the requirement of being power-counting renormalizable \cite{Horava:2008ih,Horava:2009uw}. Since we want to keep the function $N(x)$  as a Lagrange multiplier for the Hamiltonian constraints, \ie $N$ must appear linearly in the Hamiltonian, no integrations by parts are allowed within the constraints (because $N$ depends on space) if we are to retain the canonical form for $H$ (\ref{ham}). This entails a proliferation of the number of terms that are allowed\footnote{Effects of terms of the type $N^{-1}\nabla_i N$ to remedy some of the aspects of Ho\v{r}ava gravity have been studied for example in \cite{Blas:2009qj}.} but we have not written them all in (\ref{H2}) since our results will not depend on the details of these terms.
Note that standard general relativity with Minkowskian signature can be described in using the equations above and corresponds to the choice $\lambda = 1$, $\xi <0$, $\sigma$ arbitrary and all the other parameters equal to zero.

Contrary to what happens in standard general relativity the lapse $N$ is not arbitrary in Ho\v{r}ava gravity because the theory is not invariant under all spacetime diffeomorphisms. An important requirement, however, is that although $N$ is not arbitrary, there is enough freedom in $N$ so that there exist acceptable $N$'s which do not vanish, $N \not= 0$.   Otherwise, if the only acceptable $N$'s are zero, there is no true dynamics.  By contrast, the shift $N^k$ is arbitrary, reflecting full spatial diffeomorphism invariance. In the gauge $N^k=0$, which we will use, the Hamiltonian reduces to $H = \int \text{d}^3x \, N(x) \, {\mathcal H}(x)$.

\subsection{Condition on the lapse}
The equations of motion are given by
\begin{equation}
\dot{F} = [F,H]
\end{equation}
for any function(al) $F[g_{ij}(x),\pi^{ij}(x)]$ of the canonical variables, together with the above constraints.  The momentum constraints, which generate a gauge symmetry (spatial diffeomorphisms) are first-class.  However, the Hamiltonian constraints are not first-class in Ho\v{r}ava gravity and as we will see below, their preservation in time is more subtle as we now discuss.

Requesting that the constraint surface be preserved by the dynamics, \ie $\dot{{\mathcal H}}(x) = [{\mathcal H}(x), H] =  \int \text{d}^3 y \, G(x,y) N(y) \approx 0$, with $G(x,y) = [ {\mathcal H}(x),  {\mathcal H}(y)] $, leads to a partial differential equation (in space) for the lapse function of the form \begin{equation}
\alpha^{ijkl} \nabla_{ijkl} N + \beta^{ijk} \nabla_{ijk} N + \gamma^{ij} \nabla_{ij} N + \delta^{i} \nabla_{i} N + \omega N\approx 0  \label{key}
\end{equation}
where $\alpha^{ijkl} = \alpha^{(ijkl)} $, $ \beta^{ijk} =  \beta^{(ijk)}$, $\gamma^{ij} = \gamma^{ji}$, $\delta^i$ and $ \omega$ are functions of the canonical variables that depend on the coupling constants and $\nabla_{ij} =\nabla_{(i}\nabla_{j)}$ etc. The explicit form of the coefficients will not be needed here. We shall only need two crucial facts: (i) The coefficient of a given coupling constant in $\beta^{ijk}$ contains one more derivative than the corresponding coefficient in $\alpha^{ijkl}$, that in $\gamma^{ij}$ contains two more derivatives, etc. This just follows from dimensional analysis. Thus if the coefficient of one coupling constant in $\alpha^{ijkl}$ generically goes like $\frac{1}{r^a}$ at infinity (in the asymptotically flat case), the corresponding coefficient in $\beta^{ijk}$ will generically go like $\frac{1}{r^{a+1}}$, that in $\gamma^{ij}$ will go like $\frac{1}{r^{a+2}}$, that in $\delta^{i}$ will go like $\frac{1}{r^{a+3}}$  and that in $\omega$ will go like $\frac{1}{r^{a+4}}$.  (ii) Generically $\omega$ does not vanish, $\omega \not= 0$, not even weakly. This will be explicitly verified below.

In the case of general relativity, the functions $\alpha^{ijkl}$, $ \beta^{ijk}$, $\gamma^{ij}$, $\delta^i$ and $ \omega$ are all zero on-shell and the equation (\ref{key}) puts therefore no restriction on $N$ (reflecting full time reparametrization invariance). This also occurs for Euclidean general relativity which has the same values of the parameters as Minkowskian relativity except that $\xi$ is positive, $\xi >0$, as well as for zero Hamiltonian signature spacetimes \cite{Teitelboim:1981fb,Isham:1975ur,Henneaux:1979vn}, which have instead $\xi=0$. For generic values of the coupling constants, the functions $\alpha^{ijkl}$, $ \beta^{ijk}$, $\gamma^{ij}$, $\delta^i$ and $ \omega$ do not vanish and the condition (\ref{key}), which expresses that $\int \text{d}^3 x N(x) {\mathcal H}(x)$ is first-class, is then non-trivial\footnote{The importance of the condition (\ref{key}) and the fact that it is non trivial have been pointed out earlier in \cite{Kocharyan:2009te}.  However,  as far as we can see, this interesting work does not provide a detailed analysis of that equation (which always possesses the solution $N=0$).  We thank A. A. Kocharyan for pointing out his work to us.}.

In the case of closed spatial sections, there is no further condition on the lapse and Eq. (\ref{key}) is everything. In the case of open spatial sections, the lapse should obey additional boundary conditions at infinity expressing that the motion defines an asymptotic time translation.  {}For asymptotically flat spaces, which we shall consider from now on (we therefore set $\sigma = 0$), this means
\begin{equation} N \rightarrow C  \; \; \hbox{ for $r \rightarrow \infty$},
\end{equation}
where $C$ is a constant.
[There are then also additional surface terms at spatial infinity in the expressions for the generators.]

The reason for which the equation (\ref{key}) is rather complicated to analyse is that the rank of the kernel $G(x,y)$ is not constant on the constraint surface defined by the Hamiltonian and momentum constraints.  For instance, for static hypersurfaces (vanishing extrinsic curvature, $K_{ij} = 0$) -- a case much studied in the literature \cite{Lu:2009em,Kiritsis:2009rx,Kiritsis:2009vz} --, $G(x,y)$ is zero.   Similarly, if the spatial sections are of constant curvature and the extrinsic curvature is a (time-dependent) multiple of the metric, as it is relevant for cosmological models \cite{Calcagni:2009ar,Kiritsis:2009sh,Bakas:2009ku}, the covariant derivatives of the spatial Riemann tensor and of the extrinsic curvature vanish so that $G(x,y)$ is also zero. In those instances, Eq.(\ref{key}) completely degenerates ($0=0$) and brings no restriction on $N$. These special cases of measure zero are blind to the restrictions derived from (\ref{key}).  However, this is not the case for generic configurations as we shall analyse in more detail in the next section. It should be stressed that the constraints (\ref{key}) on the lapse have a fundamental character and are different from the constraints that one gets by imposing a particular ansatz on the fields.  For instance, if one imposes staticity (zero extrinsic curvature), the preservation in time of the equation $K_{ij} = 0$ leads to equations on the lapse.  But contrary to the constraints (\ref{key}), which should be fulfilled by {\em all} solutions, these particular equations are less fundamental as they depend on the ansatz.

Systems for which the rank of the brackets of the constraints is not constant on the constraint surface were discarded by Dirac \cite{Dirac:1950pj,Henneaux:1992ig} in his analysis of constrained systems.  This is because the change in the rank is somewhat pathological as one cannot globally define the Dirac bracket so that it is not clear how to consistently quantize such systems.  One can nevertheless try to apply Dirac methods in regions of the constraint surface where the rank is constant.  Although there were no known physical models exhibiting such a phenomenon at the time of \cite{Dirac:1950pj}, examples have been encountered more recently \cite{Banados:1995mq,Banados:1996yj,Banados:1997qs}.  The regions where the rank achieves its highest value compatible with the constraints are open subsets of the constraint surface and correspond to the generic situation.  Regions where the rank achieves smaller values are defined by equations, namely, precisely the equations expressing that the rank has a value smaller than the highest one.  These regions have thus smaller dimensionality and define ``non-generic" situations.

\section{Analysis of the equation on the lapse}
\label{sec:lapse}

Before analysing in more detail the equation (\ref{key}) for $N$, let us gain more insight into the model by studying its gauge invariances.

\subsection{Gauge invariances of the action}

The action (\ref{action0}) is invariant under arbitrary spacetime-dependent spatial diffeomorphisms,
\begin{subequations}
\label{spacerep}
\begin{align}
\delta g_{ij} &= \eta^kg_{ij\; ,k} + \eta^k_{\;\; ,i}g_{kj} + \eta^k_{\; \; ,j}g_{ik} , \\
\delta \pi^{ij} &= (\eta^k \pi^{ij})_{ ,k} - \eta^i_{\; \; ,k} \pi^{kj} - \eta^j_{\; \; ,k} \pi^{ik}, \; \; \\
\delta N &= \eta^k N_{, k} , \\
\delta N^i &= \dot{\eta}^i + \eta^k N^i_{\;\; ,k} - \eta^i_{\; \; ,k}N^k
\end{align}
\end{subequations}
where $\eta^k(t,x)$ are the components of the arbitrary vector field defining the diffeomorphism and the comma denotes partial differentiation. It is also invariant under space-independent time reparametrizations $\eta(t)$,
\begin{subequations}
\label{timerep}
\begin{align}
\delta g_{ij} &= \eta \dot{g}_{ij}, \label{TR1}\\
\delta \pi^{ij} &= \eta \dot{\pi}^{ij}, \label{TR2}\\
\delta N &= (\eta N)\dot{} \, , \label{TR3}\\
\delta N^k &= (\eta N^k)\dot{}\, .\label{TR4}
\end{align}
\end{subequations}
There is no other independent gauge symmetry for generic values of the coupling constants.

To the spatial diffeomorphisms correspond the first-class constraints ${\mathcal H}_k \approx 0$, as we already pointed out. This is in agreement with the general rule that``gauge invariances are generated by first-class constraints" \cite{Dirac:1950pj,Henneaux:1992ig}.  One might therefore conjecture on this basis that to the time reparametrization invariance of the action should also be associated one first-class constraint, and that there should be no other first-class constraints for generic values of the coupling constants (a conclusion that we will show to be wrong, but for now let us stick to that standard logic).  Now, the gauge symmetries in Hamiltonian form are generated by combinations of the constraints of the form $$H[\xi, \xi^k] = \int \text{d}^3 x \left( \xi(x) \, {\mathcal H}(x) +  \xi^k(x) \, {\mathcal H}_k(x)\right)$$ where $\xi$ should obey the same equation
\begin{equation}  \alpha^{ijkl} \nabla_{ijkl} \xi + \beta^{ijk} \nabla_{ijk} \xi + \gamma^{ij} \nabla_{ij} \xi + \delta^{i} \nabla_{i} \xi + \omega \xi \approx 0  \label{key3} \end{equation} as the lapse (see appendix A).

Since $$ \frac{\partial}{\partial t} = N \, n + N^k  \frac{\partial}{\partial x^k}$$ where $n$ is the unit normal to the hypersurfaces $t = const$, one finds that the component $\xi$ along the normal of the vector field $$\eta^\mu = \eta(t) \frac{\partial}{\partial t} + \eta^k(t,x)  \frac{\partial}{\partial x^k}$$ is given by \begin{equation}\label{eta}\xi = \eta N\end{equation} where the space dependence of $\xi$ occurs only through the lapse, the function $\eta$ depending only on time (and thus being a constant at any given time).

Furthermore, because the equation (\ref{key3}) for $\xi$ is linear homogeneous in $\xi$, one finds that if $\xi_0$ is a solution, then any multiple of $\xi_0$ is also a solution. Thus, in view of (\ref{eta}) and the fact that the equation for $\xi$ is the same as the equation for $N$, given a particular solution $N$, $\eta (t)N$ is also a solution for any $\eta$.  This means that a non-trivial time-reparametrization invariance is guaranteed to hold, {\it provided there exists a non-trivial solution of (\ref{key}) for N}.  In addition, since we do not expect other invariances besides the diffeomorphisms described above, all the solutions of (\ref{key3}) should be of the form $\eta N$, \ie all the solutions should be multiples of the lapse.

\subsection{Three conjectures on the rank of $G(x,y)$ (two false, one right)}

In view of the above discussion, one might be led to conjecture:

\vspace{.2cm}
\noindent
{\bf Conjecture 1:} {\em The corank of $G(x,y)$ is equal to one, i.e., the space of (admissible) solutions of the homogeneous equation (\ref{key}) or (\ref{key3}) is one-dimensional.  All solutions are multiples of a given, non-vanishing solution}.\footnote{Here, ``admissible" means ``sufficiently smooth and obeying appropriate boundary conditions at infinity in the case of open spatial sections".}

\vspace{.2cm}

In the asymptotically flat case, the time reparametrizations are true (``proper") gauge symmetries only if they vanish at infinity \cite{RT,BCT}.  However, since in Horava gravity the time reparametrizations do not depend on space, they are expected to vanish everywhere if they vanish at infinity and we might again want to conjecture:

\vspace{.2cm}
\noindent
{\bf Conjecture 2:} {\em In the asymptotically flat case, the only solution of (\ref{key}) or (\ref{key3}) that vanishes at infinity is zero everywhere.  Put differently,  the corank of $G(x,y)$ in the space of functions that vanish at infinity is zero (and thus $G(x,y)$ is formally invertible in that space).}

\vspace{.2cm}

The time reparametrizations that go to a non-vanishing constant at infinity are not be be thought of as true gauge tranformations but rather as rigid symmetries. They are sometimes called ``improper gauge transformations" \cite{BCT}. This is because they are generated by a combination of the constraints plus an appropriate, non-vanishing, surface term at infinity that makes the functional derivatives of the generator well-defined \cite{RT}.\footnote{In addition, in order to preserve the boundary conditions, they must fulfill $\dot{\eta} = 0$ at infinity and thus everywhere; only time-independent time translations are full symmetries among the time reparametrizations.}

Since the difference between two solutions of (\ref{key}) or (\ref{key3}) that go to the same constant at infinity is a solution of (\ref{key}) or (\ref{key3}) that goes to zero at infinity, one might be tempted to formulate the third conjecture:

\vspace{.2cm}
\noindent
{\bf Conjecture 3:} {\em In the asymptotically flat case, there is only one solution of (\ref{key}) or (\ref{key3}) that goes to a given constant at infinity. Therefore, the space of solutions of (\ref{key}) or (\ref{key3}) that go to an arbitrary given constant at infinity is one-dimensional. That is, the corank of $G(x,y)$ is equal to one in that space and all solutions are multiples of a given, non-vanishing solution, which may be assumed to go to one at infinity (the standard lapse).}

\vspace{.2cm}

We shall explicitly establish that, contrary to expectations, the third conjecture is incorrect for generic values of the coupling constants.  More precisely, there is no solution that goes to a non-vanishing constant at infinity for generic configurations $g_{ij}$, $\pi^{ij}$ that solve the constraints and obey the standard asymptotically flat space conditions (see below).  If the second conjecture is correct (and we believe it is), this indicates that the corank of $G(x,y)$ is equal to zero also in the enlarged space of functions allowed to go to a non-vanishing constant at infinity.  This would thus invalidate conjecture 1.  In fact, we shall be able to explicitly prove the second conjecture for a specific choice of the coupling constants that make the equations more tractable.  Thus conjecture 1 is evidently incorrect in that case and the only solution of the equation (\ref{key}) for the lapse is $N=0$, making the Hamiltonian trivial and quite different from that of general relativity.   We shall also argue, using a genericity argument, that this property remains true for generic values of the coupling constants.  In section \ref{sec:paradox}, we shall explain why there is no contradiction with the well-established rule that gauge invariances imply first-class constraints, because reparametrization in time turns out to be an on-shell trivial gauge symmetry.

Before proceeding with the analysis of the conjectures in the asymptotically flat case, we first clarify another seeming paradox that might come to mind if one compares the non-projectable class of theories with the projectable one.  This will also shed some crucial light on the form of the $\omega$ coefficient in (\ref{key}) or (\ref{key3}).

\subsection{Comparison with the projectable case}\label{proj}

The fact that the projectable theory where the lapse depends only on time is invariant under time reparametrization might suggest that  Eq. (\ref{key3}) always possesses one solution, namely, $\xi$ independent of the spatial coordinates.  This would be the case if and only if the coefficient $\omega$ of the undifferentiated $\xi$ in Eq. (\ref{key3}) were equal to zero, $\omega = 0$.  However, this is generically not the case and $\xi = const$ is therefore not a solution.

The reason for which there is no contradiction is the following. The projectable case is described by the action
\begin{equation}
S[g_{ij}(x,t),\pi^{ij}(x,t),N(t),N^k(x,t)] = \int \text{d}t \left[\left(\int \text{d}^3x \, \pi^{ij} \dot{g}_{ij}\right) - H^{\hbox{{\tiny proj}}} \right]  \label{projectable}
\end{equation}
with $ H^{\hbox{{\tiny proj}}}$ given by
\begin{equation}
 H^{\hbox{{\tiny proj}}} = N \int \text{d}^3x \, {\mathcal H} \, + \int \text{d}^3x \, N^k {\mathcal H}_k
\end{equation}
In that case, invariance of the action does not require the local condition $\omega = 0$,  but only its integrated version (see Appendix A),
\begin{equation}
\int \text{d}^3x \, \omega = 0. \label{integrated}
\end{equation}
When (\ref{integrated}) holds, then the action (\ref{projectable}) is invariant under the transformation generated by $\xi \int \text{d}^3x \, {\mathcal H} \, + \int \text{d}^3x\, \xi^k {\mathcal H}_k$ provided one transforms at the same time the lapse and the shift as
$$\delta N = \dot{\xi}, \; \; \delta N^k = \dot{\xi}^k + \xi^m N^k_{,m} - N^m \xi^k_{,m}.$$

Now, while $\omega \not= 0$ in the generic case, its integral vanishes (for any choice of ${\mathcal H}$) by virtue of the antisymmetry of the Poisson bracket.  Indeed, $[{\mathcal H}(x), {\mathcal H}(y)] = - [{\mathcal H}(y), {\mathcal H}(x)]$ implies $$\omega = - \omega +\partial_k V^k \;\;\text{for some}\;\; V^k,\; \text{\ie}\; \omega = \partial_k ( V^k/2),$$ which leads to (\ref{integrated}).  The same argument would hold if instead of $N= N(t)$ one would take the lapse to be of the form $N= \bar{N}(t) f(x)$ for some fixed function $f$ of the spatial coordinates, and the gauge transformations to be generated by $$\bar{\xi} \int \text{d}^3x \, f {\mathcal H} \, + \int \text{d}^3x\, \xi^k {\mathcal H}_k \;\;\text{with}\;\; \bar{\xi} = \bar{\xi}(t).$$

Note that, incidentally, it follows more generally from the antisymmetry argument that $\xi (\alpha^{ijk} \nabla_{ijk} \xi + \beta^{ij} \nabla_{ij} \xi + \gamma^i \nabla_i \xi + \omega \xi )$ is a spatial divergence and so (\ref{key2}) can be rewritten in the form $\nabla_k (M^k) \approx 0$ for some $M^k$ quadratic in $\xi$ and its derivatives.

\subsection{Asymptotically flat spaces -- behaviour of the fields at spatial infinity}
Since general relativity is supposed to be recovered at large distances, we shall impose the familiar asymptotic behaviour of Einstein gravity on the metric, the extrinsic curvature, the lapse and the shift. These are (in asymptotically flat coordinates \cite{RT}):
\begin{align}
g_{ij} &= \delta_{ij} + O\left(\frac{1}{r}\right), & \pi^{ij} &= O\left(\frac{1}{r^2}\right), \\
N &= 1 + O\left(\frac{1}{r}\right), & N^k &= O\left(\frac{1}{r}\right).
\end{align}
One should also impose appropriate parity conditions on the leading orders of the deviation from Minkowski space, but these will not be explicitly written here as they are not relevant to the discussion.  Note that most of the solutions with $\sigma = 0$ given in \cite{Lu:2009em,Kiritsis:2009rx,Kiritsis:2009vz} obey these boundary conditions.

\subsection{Conjecture 3 is false}

We now turn to the equation (\ref{key3}) for $\xi$, requesting the behaviour $ \xi = C + O\left(\frac{1}{r}\right)$ at infinity (the lapse corresponds to the constant $C$ taken equal to one).  One gets $$\nabla_i \xi = O\left(\frac{1}{r^2}\right),\;\; \nabla_{ij} \xi = O\left(\frac{1}{r^3}\right),\;\; \nabla_{ijk} \xi = O\left(\frac{1}{r^4}\right)\;\; \text{and}\;\; \nabla_{ijkl} \xi = O\left(\frac{1}{r^5}\right)$$ where $\nabla_i \xi$ is two orders below the leading order of $\xi$ because $C_{,i} = 0$. Given this behaviour, we find that the only leading order term in the equation (\ref{key3}) comes from the part involving $\omega$ and is given by
$$ \omega_{\text{leading}} C.$$
If the leading term of $\omega$ does not vanish, imposing the equation (\ref{key3}) forces $C$ to vanish.  Hence, there is no solution to (\ref{key3}) that tends to a nonvanishing constant at infinity.

This result crucially depends on the fact that the leading term of $\omega$ is not zero.  One might argue that because the coefficients $\alpha^{ijkl}$, $ \beta^{ijk}$, $\gamma^{ij}$, $\delta^i$ and $ \omega$ are not independent (they are functions of the canonical variables, which are furthermore constrained by the Hamiltonian and momentum constraints), it might happen that the leading term of $\omega$ vanishes when this dependence is taken into account.  We shall explicitly verify below that this ``miracle" does not occur for a particular choice of the couplings and for generic values of the canonical variables fulfilling the constraints.  By continuity, it does not occur for neighbourhing values of the couplings (the property $\omega_{\text{leading}} \not=0$ is an inequality).

\section{A tractable choice of coupling constants}
\label{sec:tractable}

Because the equation (\ref{key}) for $N$ is rather intricate, we now turn to analyse it in the case of a simpler model that was also studied in \cite{Blas:2009yd}.

\subsection{The model}

The model \cite{Blas:2009yd} is obtained by setting all couplings equal to zero, except $\lambda$ and $\xi$. In order to depart from general relativity, we take $\lambda \not= 1$. Defining $u = N^2$, the equation (\ref{key}) for the lapse then reduces to
\begin{equation}
\nabla_i(u \nabla^i \pi) = 0 \label{key4}
\end{equation}
which can be rewritten as
\begin{equation}
\nabla^i \pi \nabla_i u + \triangle \pi u = 0 \label{key4'}
\end{equation}
We note the asymptotic behaviours (again in asymptotically flat coordinates)$$\nabla^i \pi = O\left(\frac{1}{r^3}\right), \; \; \; \nabla_i u = O\left(\frac{1}{r^2}\right),  \; \; \; \triangle \pi = O\left(\frac{1}{r^4}\right), \; \; \; u = O(1).$$

The problem is now to determine the general solution of (\ref{key4}) that goes to a constant at infinity, given that the metric and its conjugate momentum are subject to the momentum and Hamiltonian constraints and to the boundary conditions.  We consider generic metric and conjugate momentum configurations compatible with the constraints and the boundary conditions, and not particular configurations subject to additional restrictions (of symmetry nature or of a different type).

What enables one to proceed in this case are results on the solutions of the constraint equations established long ago for general relativity in \cite{York:1971hw,O'Murchadha:1974nc,O'Murchadha:1974nd}.  What these authors show is that one can freely choose the trace $\pi$ of the conjugate momentum and its traceless transverse part.  The momentum constraints determine then its longitudinal (or ``vector") part.  Similarly, one can freely specify the spatial metric up to a conformal factor, which is fixed by the the Hamiltonian constraints (we refer to the original works for the details).  In these works, the Hamiltonian constraints are those of general relativity ($\lambda =  1$) but the equation for the conformal factor keeps the same form if one changes $\lambda$, thus allowing us to analyse (\ref{key4}) assuming that $\pi$ and the spatial metric (up to the conformal factor) are completely independent and unconstrained.\footnote{\label{fn:stronger}Actually, the authors of \cite{O'Murchadha:1974nc,O'Murchadha:1974nd} assume that $\pi$ decreases slightly faster than $O\left(\frac{1}{r^2}\right)$, namely, $ \pi = O\left(\frac{1}{r^{2+\epsilon}}\right)$ with $\epsilon$ arbitrarily small but strictly positive.  This makes $\triangle \pi = O\left(\frac{1}{r^{4+\epsilon}}\right)$, which still dominates the first $O\left(\frac{1}{r^5}\right)$ term (which is actually $O\left(\frac{1}{r^{5+ \epsilon}}\right)$ in (\ref{key4})).}

\subsection{Analysis of equation (\ref{key4})}

In order to further analyse the equation (\ref{key4}), let us rewrite it in terms of $K=g^{ij}K_{ij}$. The advantage of this is that $K$ is a scalar, making the transition from Cartesian to polar coordinates easier. The equation (\ref{key4}) then becomes $\partial_i (u\sqrt{g} \nabla^iK) = 0$. Now, similarly to $\pi$, $K$ can be chosen at will, so that the coefficient $\omega \sim \triangle K$ of the undifferentiated lapse in (\ref{key})  does not generically vanish on the constraint surface, in agreement with our claim above.

Using equation (\ref{key4}) written in terms of $K$, we now prove that its solutions of generically blow up at infinity.  To that end, we first consider a configuration in which $K$ does not vanish and depends only on $r$, while the metric is diagonal. In polar coordinates, (\ref{key4}) then becomes
$$\partial_r(u \sqrt{g} \nabla^r K) = 0$$
whose solution is $u = \frac{\al(\theta, \phi)}{\sqrt{g} \nabla^r K} $. Since $ \sqrt{g} \nabla^r K $ goes to zero in polar coordinates like $O\left(\frac{1}{r}\right)$\footnote{Even like $O\left(\frac{1}{r^{1+\epsilon}}\right)$ if one adopts the stronger boundary conditions mentioned in footnote~\ref{fn:stronger}.}, we conclude that $u$ blows up at infinity unless $\al=0$, in which case $u$ vanishes everywhere.  Thus, the only solution that goes to a constant at infinity vanishes identically.\footnote{Note that if $\al \not = 0$, there is also a singularity at the origin $r=0$.}  We show in appendix B that this result remains valid when we no longer impose any restriction on $K$ and $g_{ij}$.

The general conclusion is then that the only solution of (\ref{key4}) that goes to a constant at infinity is the identically vanishing solution.  In particular, the only solution that tends to zero at infinity is then generically the solution $u=0$, which establishes conjecture 2 for the above choice of coupling constants.

Note again the importance of the word ``generically" in this sentence. Indeed, there exist specific configurations of the metric and of the extrinsic curvature compatible with the constraints such that $u \not= 0$, \eg $K = 0$, but these only represent a particularization of the generic case we want to study (if one considers $K=0$, just perturb generically $K$ away from zero, which is permissible on the constraint surface according to \cite{O'Murchadha:1974nc,O'Murchadha:1974nd}).

\subsection{Mathematical consistency versus physical considerations}

The fact that the Hamiltonian constraints are second-class is not, in itself, a mathematical inconsistency.  It simply tells us that the lapse is uniquely fixed, and, because the equation for $N$ is a homogeneous equation always possessing $N=0$ as a solution, it means that $N = 0$.  The theory then possesses $5/2$ degrees of freedom per space point since there are $6$ conjugate pairs, $3$ first-class constraints and one second-class constraint (per space point). We thus agree with reference \cite{Blas:2009yd}, which also concluded that the Hamiltonian constraints were (generically) second-class and determined the lapse.  However, we go beyond this work by proving more completely that the constraints are indeed second-class and drawing the inevitable conclusion that then, necessarily, $N=0$ (homogeneity of the equation for $N$).

The extra $1/2$ degree of freedom (the so-called ``extra mode") might be thought of as contained in the pair formed by $\pi$ and the conformal factor (and not in $N$, which is identically zero).  The conformal factor is determined by the Hamiltonian constraints.  In general relativity, where the constraints are first-class, one uses the corresponding gauge freedom to impose a gauge condition on the conjugated $\pi(x)$. Here, the constraints are second-class, thus expressing instead that $\pi(x)$ is self-conjugate in the corresponding Dirac bracket (whose expression is rather intricate and will not be worked out here). We note that the extra mode is somewhat analogous to a chiral boson \cite{Floreanini:1987as,Henneaux:1988gg}, for which there is also a single second-class constraint per space point.

Since $N = 0$, the dynamics is very simple: the Hamiltonian vanishes (in the gauge where the shift is zero) and any function of the canonical variables is a constant of motion.  This is mathematically consistent but the theory not only differs in a drastic way from general relativity but is also physically rather meaningless as there is no time evolution.  One can therefore say that there is a dynamical inconsistency with what one requests from the theory on physical grounds, \ie the lapse should be non zero and belong to a one-parameter family of solutions (away from the general relativity values).

\subsection{Extension to other values of the couplings by continuity arguments}
We have shown that the rank of the ``matrix" $[{\mathcal H}(x), {\mathcal H}(x')]$ is (generically) maximum for the specific values of the couplings considered in this section.  What about other values of the couplings? If the matrix under consideration were a finite $\text{N} \times \text{N}$ matrix, one could use continuity arguments to argue that it has also the maximum rank N for neighbouring values of the couplings.  Indeed, if a matrix $\text{M}(\lambda_i)$ depending continuously on a set of parameters is invertible for some specific values $\{\lambda_i^0\}$  of these parameters, then it is generically invertible by continuity (the equation $\det \text{M}(\lambda_i)= 0$ is a submanifold of lower dimension in parameter space). More generally, the highest value of the rank achieved by M, which might be smaller than N if the matrix is nowhere invertible, is in that sense generic.  One may invoke the same arguments here, but they should be taken with a grain of salt since we are dealing with infinite matrices.

The fact that conjecture 3 has been explicitly shown to be incorrect, and that conjecture 2 seems plausible in the philosophy of Ho\v{r}ava theory, gives further support to the belief that {\em all} the Hamiltonian constraints are second-class for generic values of the coupling constants and not just for those considered here.

\section{Reparametrization invariance as an on-shell trivial symmetry}
\label{sec:paradox}

\subsection{A paradox?}
As we stated, it might seem paradoxical that all the Hamiltonian constraints (\ref{hamconstraints}) are second-class, while the action is invariant under time reparametrizations as in (\ref{timerep}). This invariance would seem to imply that at least one combination of the Hamiltonian constraints (\ref{hamconstraints}) should be first-class and generate the time reparametrizations rewritten in Hamiltonian form. This combination would necessarily correspond to a non-trivial solution $N$ for the key equation (\ref{key}).

However, the point is that the time reparametrizations (\ref{timerep}) are ``on-shell trivial" gauge symmetries when the Hamiltonian constraints are all second-class and therefore have no non-trivial physical content and need not be associated with first-class constraints among the Hamiltonian constraints, thus implying that a non-trivial solution to the equation (\ref{key}) is not needed (so that no contradiction arises).

We will show this below on a simpler example but, first, let us say a word about these so-called ``on-shell trivial" gauge symmetries. It is well known that any action $S[y^A]$ is always invariant under the transformations
\begin{equation}
\delta y^A = \epsilon^{AB} \frac{\delta S}{\delta y^B} , \; \; \; \; \epsilon^{AB} = - \epsilon^{BA} \label{trivial}
\end{equation}
where $\epsilon^{AB}$ are arbitrary spacetime functions and $A, B$ run over all canonical variables.  These fake gauge transformations, which vanish on-shell, are always present and have no implication on the (classical or quantum) dynamics of the theory (see \eg \cite{MH90,Henneaux:1992ig}).  They have been studied at length in the BRST approach to the quantization of gauge systems with an ``open algebra" \cite{Kallosh:1978de,deWit:1978cd,Batalin:1981jr,Batalin:1984jr,Batalin:1984ss,Henneaux:1992ig}.  They do not need any ghost and have been shown to be associated with canonical transformations in the antibracket (see \cite{Voronov:1982cp}).

\subsection{A simpler example}

To illustrate the point that the time reparametrizations (\ref{timerep})) are ``on-shell trivial" gauge symmetries, we consider a model with the same features but with a finite number of degrees of freedom, whose action reads
\begin{equation} S[q^i, p_i, N^\alpha] = \int \text{d}t \left( p_i \dot{q}^i - N^\alpha {\mathcal H}_\alpha \right).  \label{ClassMechModel}
\end{equation}
We assume that the constraints are all second-class, so that defining \linebreak $[{\mathcal H}_\alpha, {\mathcal H}_\beta] \equiv C_{\alpha \beta}$ we have $\det(C_{\alpha \beta}) \not= 0$, and denote the inverse matrix by $C^{\alpha \beta}$, so that $C^{\rho \alpha} C_{\alpha \beta} = \delta^\rho_\alpha $.	

The multipliers $N^\alpha$ of this simpler model are to be thought of as being the analog of the lapse $N(x)$ in Horava's theory. There is no analog of the spatial diffeomorphisms here and hence no analog of the shift.

The action (\ref{ClassMechModel}) is clearly invariant under the time reparametrizations
\begin{align}
 \delta q^i &= \eta \dot{q}^i, & \delta p_i &= \eta \dot{p}_i, &\\
 \delta N^\alpha &= (\eta N^\alpha)\dot{}, & \delta {\mathcal H}_\alpha &= \eta\dot{\mathcal H}_\alpha,
\end{align}
but these transformations, which vanish when the equations of motion are satisfied, are on-shell trivial as they can identically be rewritten as antisymmetric combinations of the equations of motion as
\begin{eqnarray}
&& \delta q^i = \eta \left[ \frac{\delta S}{\delta p_i} + \frac{\partial {\mathcal H}_\rho}{\partial p_i} C^{\rho \alpha} \left(- \frac{\text{d}}{\text{d}t} \frac{\delta S}{\delta N^\alpha} - \frac{\partial {\mathcal H}_\alpha}{\partial q^j} \frac{\delta S}{\delta p_j} + \frac{\partial {\mathcal H}_\alpha}{\partial p_j} \frac{\delta S}{\delta q^j} \right) \right] \\
&& \delta p_i = \eta \left[ - \frac{\delta S}{\delta q^i} - \frac{\partial {\mathcal H}_\rho}{\partial q^i} C^{\rho \alpha} \left(- \frac{\text{d}}{\text{d}t} \frac{\delta S}{\delta N^\alpha} - \frac{\partial {\mathcal H}_\alpha}{\partial q^j} \frac{\delta S}{\delta p_j} + \frac{\partial {\mathcal H}_\alpha}{\partial p_j} \frac{\delta S}{\delta q^j} \right) \right] \\
&& \delta N^\alpha = \frac{\text{d}}{\text{d}t}\left[ \eta C^{\rho \alpha} \left(- \frac{\text{d}}{\text{d}t} \frac{\delta S}{\delta N^\alpha} - \frac{\partial {\mathcal H}_\alpha}{\partial q^j} \frac{\delta S}{\delta p_j} + \frac{\partial {\mathcal H}_\alpha}{\partial p_j} \frac{\delta S}{\delta q^j} \right) \right]
\end{eqnarray}
where we have used
\begin{equation}
N^\rho = C^{\rho \alpha} \left(- \frac{\text{d}}{\text{d}t} \frac{\delta S}{\delta N^\alpha} - \frac{\partial {\mathcal H}_\alpha}{\partial q^j} \frac{\delta S}{\delta p_j} + \frac{\partial {\mathcal H}_\alpha}{\partial p_j} \frac{\delta S}{\delta q^j} \right) .
\end{equation}
We note that the appearance of the inverse $C^{\alpha\beta}$ clearly signals that this argument only holds if all the constraints are second-class. But this is generically the case in Ho\v{r}ava gravity so that there is no non-trivial time-reparametrization even thought the action is invariant under (\ref{timerep}) (which is said to be ``on-shell trivial").

\section{Conclusions}

In this paper, we have uncovered several problems in the non-projectable class of theories considered in \cite{Horava:2008ih,Horava:2009uw}.  Besides the fact that these theories do not fulfill the standard regularity condition on the rank of the ``matrix" of the Poisson brackets of the constraints, we have generically shown that they do not admit solutions for the lapse that go to a non-vanishing constant at infinity, \ie the lapse must asymptotically go to zero, thus preventing any asymptotic dynamics. Particular solutions, corresponding to points on the constraint surface where the rank of the ``matrix" of the Poisson brackets of the constraints is not maximum, fail to reveal this problem.

We have then considered a particular choice of the coupling constants (which is tractable and which is believed to be representative of the general situation) and have shown that the lapse must vanish in that case everywhere and not just at infinity (again, for generic solutions of the constraint equations).  There is no contradiction with time reparametrization invariance, because this invariance then turns out to be an ``on-shell trivial" gauge symmetry with no physical implication.

In order to avoid such serious difficulties, one might try to take advantage of the non-constancy of the rank of the ``matrix" $[{\mathcal H}(x), {\mathcal H}(x')]$ of the Poisson brackets of the constraints and force the system, by additional constraints, to be on a subset of the constraint surface where the equation for the lapse admits non-trivial solutions (\ie one forces the system to be on non-generic subsets).  This is the approach explored in \cite{Li:2009bg} (which, however, did not recognize the fact that generically the constraints are second-class). One such extra condition might be $K_{ij} = 0$ as this makes $[{\mathcal H}(x), {\mathcal H}(x')]$ identically zero.  This choice yields a theory with non-trivial solutions having $N \not=0$ \cite{Lu:2009em,Kiritsis:2009rx,Kiritsis:2009vz} and in that sense is consistent.   However, these extra constraints, together with the constraints following from $\dot{K}_{ij}=0$, constitute a violent simplification of the theory which dramatically reduces its number of degrees of freedom and bears little resemblance with full general relativity (it is certainly not a UV {\it extension}).

Another possible choice would be $\omega(x) = 0$ since this allows $N=N(t)$. These extra constraints reduce the number of degrees of freedom to 2 per space point or even to lower values if their preservation in time yields further constraints (an analysis which appears to be rather involved to be carried out). However,  these extra constraints do not have a clear geometrical interpretation and in any case deviate from the original proposal by Ho\v{r}ava, which would lose much of its appeal.  The resulting theory would again not be a UV {\it extension} of general relativity.  It is amusing to note that the extra constraints $\omega=0$ yields the condition $\pi = 0$ for the particular values of the couplings explicitly studied above.  This may be viewed in general relativity as a gauge condition fixing the slicing \cite{Dirac:1958jc,York:1971hw,DeWitt:1967yk}. If $\pi=0$, one may set $\lambda = 1$: the resulting theory is consistent, but is just general relativity in disguise (in a gauge-fixed formulation where its geometrical content is somewhat obscure).\footnote{One might in fact take as extra consistent constraint ${\mathcal H} - {\mathcal H}^{GR} = 0$ when this constraint is an acceptable gauge condition for fixing the slicing of Einstein theory, recovering in this way a gauge fixed version of ordinary general relativity.}

Although we have not investigated the equation for the lapse in the compact case, one might anticipate that difficulties in the analysis will also arise in that case since the solutions must be globally well-defined. Locality requirement for the lapse as a function of the other variables should presumably also be imposed in order to be able to apply the methods of local quantum field theory.  This appears to be also a very restrictive condition.

\section*{Acknowledgements} Our work is partially supported by IISN - Belgium (conventions 4.4511.06 and 4.4514.08)
and by the Belgian Federal Science Policy Office through the Interuniversity Attraction Pole P6/11. AK is a Research Associate of the Fonds de la Recherche Scientifique--FNRS, Belgium.

\appendix

\section{Gauge symmetries in Hamiltonian form}

The gauge symmetries in Hamiltonian form (after time derivatives of the dynamical variables have been eliminated through the addition of trivial transformations) are the space reparametrizations, generated by
\begin{equation}H
[\xi^k] = \int \text{d}^3 x \, \xi^k(x) \, {\mathcal H}_k(x)
\end{equation}
where $\xi^k(x)$ is an arbitrary vector field (that may depend on time).  In addition, the theory is also gauge invariant under the transformations generated by
\begin{equation}
H[\xi] = \int \text{d}^3 x \, \xi(x) \, {\mathcal H}(x),
\end{equation}
if $\xi(x)$ is chosen so that $H[\xi]$ is first-class.

Indeed, one finds that the variation of the action under the transformations \begin{equation}\delta g_{ij}(x) = [g_{ij}(x), H[\xi] + H[\xi^k]], \; \; \; \delta \pi^{ij}(x) = [\pi^{ij}(x), H[\xi] + H[\xi^k]]\end{equation} is given (up to surface terms at the time boundaries) by
\begin{equation}
\delta S = \int  \text{d}t \, \text{d}^3x  \left(\dot{\xi}  {\mathcal H} +  \dot{\xi}^k  {\mathcal H}_k  - \delta N {\mathcal H} -  \delta N^k {\mathcal H}_k -  N \delta {\mathcal H} -   N^k \delta {\mathcal H}_k \right)\end{equation}
where $\delta N$ and $\delta N^k$ are the variations  of the Lagrange multipliers $N$ and $N^k$ and where $\delta {\mathcal H}$ and $\delta {\mathcal H}_k$ read
\begin{eqnarray} && \delta {\mathcal H}(x) =  \left( \int \text{d}^3 y G(x,y) \xi(y)\right) + (\xi^k  {\mathcal H})_{,k}(x) \nonumber \\
&& \delta {\mathcal H}_k (x)= \left( \xi_{,k} {\mathcal H} + ({\mathcal H}_k \xi^m)_{,m} + {\mathcal H}_m \xi^m_{,k} \right)(x).  \nonumber
\end{eqnarray}
Inserting these expressions in the action, one gets, upon setting
\begin{eqnarray} && \delta N = \dot{\xi} + \xi^k N_{,k} - \xi_{,k} N^k + \Delta N, \nonumber \\
&& \delta N^k = \dot{\xi}^k + \xi^m N^k_{,m} - \xi^k_{,m} N^m + \Delta N^k,  \nonumber  \end{eqnarray} that the variation $\delta S$ reduces to
\begin{equation} \delta S = - \int \text{d}t \text{d}^3x \left(\Delta N  {\mathcal H} + \Delta N^k  {\mathcal H}_k + N\int \text{d}^3 y G(x,y) \xi(y) \right). \label{DeltaS} \end{equation}

The variations of the Lagrange multipliers $N$ and $N^k$ should be such that $\delta S = 0$ for any $g_{ij}$, $\pi^{ij}$, $N$ and $N^k$ (as these are freely varied in the variational principle).  Thus the expression (\ref{DeltaS}) should vanish for any $N$.  Taking the functional derivative of $\delta S = 0$ with respect to $N$, one gets the condition that $\int \text{d}^3y G(x,y) \xi(y)$ should vanish when the constraints hold, \ie
\begin{equation} \alpha^{ijkl} \nabla_{ijkl} \xi + \beta^{ijk} \nabla_{ijk} \xi + \gamma^{ij} \nabla_{ij} \xi + \delta^{i} \nabla_{i} \xi + \omega \xi \approx 0.  \label{key2} \end{equation} Conversely, when (\ref{key2}) holds, one can adjust $\Delta N$ and $\Delta N^k$ such that $\delta S = 0$.  Thus $\int \text{d}^3x \xi(x) {\mathcal H}(x)$ generates a gauge symmetry if and only if (\ref{key2}) holds.

The condition (\ref{key2}) for $\xi$ to define a gauge transformation is then, not surprisingly, exactly  the same equation as the equation (\ref{key}) that the lapse must fulfill on-shell.

\section{More detailed analysis of Eq.(\ref{key4})}

We analyse in this appendix the equation (\ref{key4}), which can be rewritten as (upon dividing by $\sqrt{g}$)
\begin{equation}
m^i \partial_i u = - \left(\nabla_i m^i \right) u  \label{key5}
\end{equation}
where $m^i$ is the vector field $ \nabla^i K$.
The equation (\ref{key5}) is a homogeneous partial differential order equation giving the variation of $u$ along the integral curves of $m^i$.  To integrate this equation, one needs to specify $u$ on a two-dimensional surface transverse to these integral curves (``initial data" for $u$).

In polar coordinates, the vector field $m^i$ and its covariant divergence are given by functions of the asymptotic form
\begin{equation}
m^r = \frac{a}{r^3}, \;\; m^\theta = \frac{b}{r^4}, \;\; m^\phi = \frac{c}{r^4}, \; \; - \nabla_i m^i =  \frac{z}{r^4} \label{FormOfM}
\end{equation}
with $a$, $b$, $c$ and $z$ functions on the $2$-sphere $\text{S}^2$ which have the same dimensionality.  There are terms decreasing faster to infinity than the terms written in (\ref{FormOfM}), but these are not relevant for the asymptotic analysis and are set to zero.

In the particular case considered in the text, one sets
\begin{equation}
a = 1, \;\;\; b=c= 0, \;\; \; z = 1
\end{equation}
and the equation can be rewritten $\partial_r (u/r)$ = 0.
We will study the equation (\ref{key5}) in an open region of ``parameter space" (\ie an open region in the space of the $m^i$'s) that contains this spherically symmetric situation.  We will prove explicitly in that case that if $u \not=0$, then $u$ blows up at infinity, so that the only non-pathological solution at infinity is $u=0$ (in the open region we shall consider). One can probably refine the argument to exhibit pathologies for a bigger range of the $m^i$'s,  but we shall not attempt to do it here as we do not feel it is worth it.  Together with the results of the previous sections, we think indeed that the point made here provides enough evidence of problems of the theory, which are believed to be insuperable if we are to retain the original physical meaning of the theory.

The restrictions we shall impose, which define our open region, are
\begin{itemize}
\item $a>0$ on $\text{S}^2$;
\item $z' \equiv \frac{z}{a} >0$ on $\text{S}^2$, implying $z>0$ on $\text{S}^2$
\end{itemize}
Since the sphere is compact, $a$ and $z'$ are bounded from below by some strictly positive number $C$, allowing us to set
$$0 < C \leq z'.$$

If $a$ does not vanish on $\text{S}^2$, the spheres are transverse to the integral curves of $m^i$.  One can thus give ``initial data" for $u$ on the sphere $\text{S}^2$ defined by $r=1$. Furthermore, if one divides by $a$ the equation (\ref{key5}) one gets as new asymptotic equation
\begin{equation}
m'^i \partial_i u = \frac{z'}{r} u  \label{key6}
\end{equation}
with $m'^i = r^3 m^i/a$, $b'=b/a$, $c'=c/a$ and thus
$$m'^r = 1, \; \; \; m'^\theta = \frac{b'}{r}, \; \; \; m'^\phi = \frac{c'}{r}, \;\;\; z' = \frac{z}{a}.$$
Since $m'^r =1$, we can take $r$ as parameter of the integral curves of the vector field, which have then the form $r = r$, $\theta = \Theta(r)$, $\phi = \Phi(r)$ with
$$ \frac{\text{d} \Theta}{\text{d}r} = m'^\theta, \; \; \; \; \frac{\text{d} \Phi}{\text{d}r} = m'^\phi.$$

Along a given integral curve (determined by the value $\theta_0$, $\phi_0$ of its intersection with the ``initial sphere" $r=1$), the function $u$ fulfills
\begin{equation}
\frac{\text{d}u}{\text{d}r} =   \frac{z'(\Theta(r), \Phi(r))}{r} u
\end{equation}
This implies that if $u$ vanishes at one point of the integral curve, then it vanishes everywhere on the integral curve (uniqueness of the solution of the first order differential equation in normal form for given initial data that can be taken anywhere along the curve).  Thus, if $u_0$ (value at $r = 1$) does not vanish initially at $\theta_0$, $\phi_0$, it vanishes nowhere on the corresponding integral curve.

Consider a point on the unit sphere with $u_0 \not=0$.  The equation for $u$ along the corresponding integral curve can be rewritten as
$$
\frac{\text{d}U}{\text{d}r} =   \frac{z'(\Theta(r), \Phi(r))}{r}
$$
with $U = \ln \vert u \vert$.  Since $z'$ is bounded from below, one has
$$ \frac{\text{d}U}{\text{d}r} \geq \frac{C}{r} $$
and by integrating from $r=1$ to $r$, one gets
$$ U \geq U_0 + C \ln r  $$
where $C>0$.  This implies that $U$, and thus also $u$ blows up at infinity ($u$ goes to $+ \infty$ if $u_0 >0$ and to $- \infty$ if $u_0 <0$).

Thus, if we want $u$ to go to a constant at infinity, we must take $u_0 = 0$ and $u$ is then equal to zero everywhere.  This is what we wanted to prove: the only solution that tends to a constant at infinity is $u = 0$ and the constant is then equal to zero.

\end{document}